\documentclass[12pt,preprint]{emulateapj}

\shorttitle{Evolution of observed Quasi Periodic Oscillations in Outburst sources}
\shortauthors{S. Mondal, S. K. Chakrabarti, and D. Debnath}

\begin{document}

\title{Is Compton cooling sufficient to explain evolution of observed QUASI PERIODIC OSCILLATIONS in Outburst sources?}

\author{Santanu Mondal~\altaffilmark{1}, Sandip K. Chakrabarti\altaffilmark{2,1}, Dipak Debnath\altaffilmark{1}} 
\altaffiltext{1}{Indian Center for Space Physics, 43 Chalantika, Garia Station Rd., Kolkata, 700084, India; santanu@csp.res.in}
\altaffiltext{2}{S. N. Bose National Centre for Basic Sciences, Salt Lake, Kolkata, 700098, India.}

\email{santanu@csp.res.in; chakraba@bose.res.in; dipak@csp.res.in;}

\begin{abstract}

In outburst sources, quasi-periodic oscillation (QPO) frequency is known to evolve in a certain way:
in the rising phase, it monotonically goes up till a soft intermediate state is achieved. In the propagating 
oscillatory shock model, oscillation of the Compton cloud is thought to cause QPOs. Thus, in order to increase 
QPO frequency, Compton cloud must collapse steadily in the rising phase. In decline phases, exactly opposite 
should be true. We investigate cause of this evolution of the Compton cloud. 
The same viscosity parameter which increases the Keplerian disk rate, also
moves the inner edge of the Keplerian component, thereby reducing the size of the Compton cloud
and reducing the cooling time scale. We show that cooling of the Compton 
cloud by inverse Comptonization is enough for it to collapse sufficiently so as to explain the QPO 
evolution. In the Two Component Advective Flow (TCAF) configuration of 
Chakrabarti-Titarchuk, centrifugal force induced shock represents boundary of the Compton cloud.
We take the rising phase of 2010 outburst of Galactic black hole candidate H~1743-322 
and find an estimation of variation of $\alpha$ parameter of the sub-Keplerian flow 
to be monotonically rising from $0.0001$ to $0.02$, well within the range suggested by magneto-rotational 
instability. We also estimate the inward velocity of the Compton cloud to be a few meters/second which is
comparable to what is found in several earlier studies of our group by empirically fitting the 
shock locations with the time of observations.

\end{abstract}


\section{Introduction}

Study of temporal variability including signatures of quasi-periodic oscillations (QPOs)
is an important aspect of astrophysics of black holes. 
Several models in the literature attempt to explain origin of low frequency QPOs. They include 
perturbation inside a Keplerian disk \citep{Trudolyubov99}, 
global disk oscillation \citep{TO00}, 
oscillation of wrapped disk \citep{Shirakawa02}, 
accretion ejection instability at the inner radius of a Keplerian disk \citep{Rodriguez00}. 
\citet{TLM98}, envisages a bounded region surrounding compact objects 
which is called the transition layer (TL) and identifies
low frequency QPOs as that associated with the viscous magneto-acoustic resonance oscillation of 
the bounded TL.
Chakrabarti and his collaborators \citep{MSC96,C04} 
showed that the oscillations of centrifugal pressure supported accretion shocks \citep{C90a} 
could cause observed low frequency QPOs. 
According to the two-component advective flow (TCAF) model \citep{CT95} 
the post-shock region itself is the Compton cloud. 
Because the shock is formed due to centrifugal force, where energy is dissipated 
and angular momentum is redistributed, post-shock region is also known as the 
CENtrifugal pressure supported BOundary Layer (CENBOL) of the black hole.
This TCAF solution has been proven to be of stable configuration (Giri \& Chakrabarti, 2013)
and Monte-Carlo simulations of spectral and timing properties through a time dependent radiative 
hydrodynamic code showed the formation of QPOs very similar to what is observed \citep{GG14}.  
The Compton cloud becomes smaller because of higher viscosity as well as higher cooling. 
Higher viscosity causes the Keplerian disk on the equatorial plane to move in. This causes the Compton
cloud to cool down. This picture is clear from the two component model of \citet{CT95} 
and the outburst source picture given in \citet{ETC96} based on it. 
To our knowledge, except TCAF, no other model is found to be capable of 
explaining continuous and simultaneous variation of spectral properties and timing properties
\citep[see,][]{DD08,DD10,DD13,DD14a,Nandi12}. 

There are mainly two reasons behind oscillation of shock wave in an accretion flow: 
i) Resonance Oscillation: when cooling time scale of the flow is comparable 
to the infall time scale \citep{MSC96}, 
this type of oscillation occurs. Such cases can be identified by the
fact that when accretion of the Keplerian disk is steadily increased, QPOs may occur in a range of 
the accretion rates, and the frequency should go up with accretion rate. Not all the QPOs may be of this type.
Some sources (for example, 2010 GX 339-4 outburst),
show signatures of sporadic QPOs during rising soft-intermediate states (where QPO frequencies
of in $\sim$6 Hz were observed for around 26 days; \citet{Nandi12}), although rise in the accretion
rates. In these cases the shock strength has to change in order that the resonance condition holds good.
ii) Non-Steady Solution: in this case, the flow has two saddle type sonic points, but Rankine-Hugoniot conditions 
which were used to study standing shocks in \citet{C89} 
are not satisfied. Examples of these oscillations are given in 
\citet{RCM97} 
where no explicit cooling was used. 
Such type of QPOs are possible at all accretion rates, outside the regime of type (i) QPOs mentioned above. QPO 
frequencies depend on viscosity (higher viscosity will remove angular momentum, bring shocks closer to the black hole,
and produce higher frequency QPOs), but not explicitly on accretion rate. 
In any case, observed QPO frequency is inversely proportional to the infall time ($t_{infall}$) in the
post-shock region. So, when low frequency (e.g., mHz to few Hz) QPOs are observed, generally during very early phase 
of an outburst or very late phase of an outburst of transient black hole candidates (BHCs), shocks are located very 
far away from black holes and size of the CENBOL is large. As a result, amount of cooling by photons 
from Keplerian disk \citep{SS73} 
is high \citep[][hereafter Paper-I]{CT95,MC13} 
and CENBOL pressure drops, moving the shock closer towards black hole \citep[][Paper-I]{MSC96,Das10,MCD14} 
until pressure (including centrifugal) is strong enough to balance the inward pull. Lower shock location increases the 
QPO frequency. Different BHCs show different oscillation frequencies during their evolution 
(both in rising and declining) 
phases. Using Propagating Oscillatory Shock (POS) model by Chakrabarti and his collaborators 
\citep{C05,C08,C09,DD10,DD13,Nandi12} 
one can satisfactorily explain origin and day-wise evolution of QPO frequencies 
during rising and declining phases of outbursting BHCs. During rising phase shock moves 
towards black holes increasing QPO frequencies monotonically with time and opposite scenario 
is observed during declining phase, mainly in hard and hard-intermediate spectral states 
of the outbursts \citep[see,][]{DD13}. 

Recently \citet{DD14a} 
showed that observed QPO frequencies can be predicted from detailed spectral analysis using Two Component Advective Flow (TCAF) model 
as a local additive table model in XSPEC. 
\citet{MDC14} and \citet{DD14b} 
also showed physical reason behind spectral state transitions from spectral model fitted parameters of TCAF model for two different 
Galactic BHCs H~1743-322, and GX~339-4 during their outbursts. Basically, the same shock location
is obtained by fitting the spectra produces QPOs through oscillations. So spectral properties are 
interlinked with timing properties as far as TCAF solution is concerned. 

In this {\it Paper}, our goal is to explain origin of observed QPO evolution from pure analytical point 
of view using Compton cooling. Biggest uncertainly being that of viscosity parameter,
we would like to have an idea of how viscosity usually vary with distance in a known source. 
We consider a transient BHC H~1743-322 during its 2010 outburst. We hope that in future, this behavior
could be used to better predict QPO evolutions. 

In August 2010, H~1743-322 was found to be active in X-rays \citep{Yamaoka10} 
with a characteristics of temporal and spectral evolutions \citep{DD13} 
similar to those observed in other transient BHCs 
\citep[see for a review,][]{RM06}. 
Detailed source description is already in literature 
\citep[][and references therein]{DD13,MDC14}. 

The {\it paper} is organized in the following way: in the next Section, we discuss 
governing equations of modified Rankine-Hugoniot (R-H) shock conditions in presence of Compton cooling. 
In \S 3, observed QPO evolution and what this tells us about viscosity variation in the disk 
as a function of radial distance. We also present phase space diagram of the 
flow in progressive days. Finally, in \S 4, we briefly discuss our results and make our concluding remarks.

\section{Shock condition and shock constant}

We assume the accreting flow to be thin, axisymmetric and rotating around vertical axis. To avoid integrating 
in a direction transverse to flow motion, we consider that the flow is in hydrostatic equilibrium in vertical
direction as in \citet{C89}.
In TCAF, CENBOL is basically the post-shock region of a low angular momentum, sub-Keplerian flow.
It is comparatively hotter, puffed up, and much like an ion supported torus \citep{Rees82}. 
Due to inverse Compton cooling effect of intercepted low energy photons from a Keplerian disk, 
energy of CENBOL decreases and is radiated away. Energy equation at the shock is modified by,
$$
\varepsilon_{+}=\varepsilon_{-} - \Delta \varepsilon,
\eqno{(1a)}
$$
where, $\Delta \varepsilon$ is the energy loss due to Comptonization.  
Baryon number conservation equation at the shock is,
$$
\dot{M_{+}}=\dot{M_{-}}.
\eqno{(1b)}
$$
Since the gas is puffed up, R-H conditions \citep{Landau59} 
have to be modified, so that only vertically integrated pressure and density parameters are important. 
This modification was first carried out in \citep{C89}, where pressure balance condition was modified using 
vertically integrated values: 
$$
W_{+} + \Sigma_{+}v_{+}^2 = W_{-} + \Sigma_{-}v_{-}^2.
\eqno{(1c)}
$$
Here, $W$ and $\Sigma$ are pressure and density, integrated in the vertical direction \citep{Matsumoto84}. 
In our solution, we use the Eq. (8a) of Paper-I as an invariant quantity at the shock, which is given by:
$$
           \frac{[M_{+}(3\gamma-1)+(\frac{2}{M_{+}})]^2}{2+(\gamma-1)M_{+}^2}=
           \frac{[M_{-}(3\gamma-1)+(\frac{2}{M_{-}})]^2}{2+(\gamma-1)M_{-}^2-\zeta},
\eqno{(2)}
$$
where, $M$, $v$ and $\gamma$ are the Mach number, radial velocity and adiabatic index of flow respectively, 
$\zeta=\frac{2 \Delta \varepsilon (\gamma-1)}{a_{-}^2}$. Here, $a$ is adiabatic sound speed.
We follow the same mathematical procedure and methodology as in Paper-I, 
to find shock location for a given cooling rate. In the standard theory 
of thin accretion flows around black holes \citep{SS73} viscosity plays a major role. \citet{GC13} 
showed formation of Keplerian disk for super-critical $\alpha$ parameter \citep{C90b}. 
Inflow angular momentum is transported outward by viscosity and allow it to fall 
into the black holes. As the shock moves closer, the angular momentum must be adjusted by
viscosity so that the shock formation is theoretically allowed. For our viscosity calculation, we use 
the relation \citep{C90b}: 
$$
\dot{M}(\lambda-\lambda^{'})=-r^{2}W_{r\phi},
\eqno{(3)}
$$
where, $W_{r\phi}=-\alpha P$, is the viscous stress, $\alpha$ being \citet{SS73}
viscosity parameter. Angular momentum variation from Eq. (3) can be written as 
$$
\Delta \lambda=\frac {\alpha r a^{2}}{v},
\eqno{(4)}
$$
where, $\Delta \lambda=(\lambda-\lambda^{'})$ is the change in angular momentum $(\lambda)$ due to viscous transport.
%

\subsection{Methodology of $\Delta \varepsilon$ calculation}

We analyze archival data of $8$ observational IDs of RXTE/PCA instrument (only PCU2, all layers) 
starting from 2010 August 9 (Modified Julian Day, i.e., MJD = 55417.2) to 2010 August 16 (MJD = 55424.1),
selected from rising phase of 2010 outburst of H~1743-322. We carry out data analysis using FTOOLS 
software package HeaSoft version HEADAS 6.14 and XSPEC version 12.8. For generation of source and 
background `.pha' files and spectral fitting (in $2.5-25$ keV energy range) using combined disk 
blackbody and power-law models, we use same method as described in \citet{DD13}. 
After achieving best fit based on reduced chi-square value ($\chi^2_{red} \sim 1$), we integrate only 
power-law component of the spectrum.
This can be written as:
$$\sum_{i=E_{l}}^{E_{u}} E_i F_{Comp}(i),$$ where, $E_{l}$ and $E_{u}$ are the lower and the upper limits of energy. 
For interstellar photo-electric absorption correction, we follow the prescription of \citet{MM83}. 
To calculate cooling time of the Compton cloud (CENBOL) from observed spectrum, we consider distance correction 
in following way: we multiply the integrated spectrum by the model normalization value ($norm$) of $\frac{4\pi D^2}{cos(i)}$, 
where `$D$' is source distance in $10$~kpc unit and `$i$' is the disk inclination angle. In case of H~1743-322, we use 
source distance $d$ = $8.5$~kpc, and $i = 75^\circ$ \citep{Steiner12}. 
We keep hydrogen column density (N$_{H}$) frozen at 1.6$\times$~10$^{22}$~atoms~cm$^{-2}$ for 
absorption model {\it wabs} and assume a $1.0$\% systematic error for all observations \citep{MDC14}. 

\section{Results}

In this {\it Paper}, we study origin and evolution of QPOs in outbursting BHC H~1743-322,
from a purely analytical point of view. In \citet{CT95} and \citet{DC04}, 
it was shown that matter from the companion is heated up due to compression and puffed up due to centrifugal barrier 
to form CENBOL. Low energy photons from a \citet{SS73} disk with an accretion rate  
of $\dot{m}_d$ are intercepted by CENBOL and are emitted as high energy photons after inverse Comptonization. 
In Fig. 1a, we show that the rate of cooling of the CENBOL in progressive days during the rising 
phase of the outburst. As day progresses, amount of cooling 
increases and shock moves towards the black hole (MSC96), which is shown in Fig. 1b. On first observed day 
of the outburst (MJD = 55417.2), location of the shock ($X_s$ in Schwarzschild radius $r_g=2GM/c^2$) 
was at $350.65~r_g$ and at the end of our observation (MJD = 55424.1), it reaches at $\sim 64.99~r_g$. 
In Fig. 1c, we show Mach number variation of the flow in $1^{st}$ day 
(solid curve, shock was at $350.65~r_g$), $5.05^{th}$ day (dashed curve, shock was at $191.69~r_g$) 
and $7.81^{th}$ day (dotted curve, shock was at $64.99~r_g$) of the outburst. We calculate the velocity 
of movement of the shock to be $\sim 13.11~ms^{-1}$, which roughly matches with the final 
velocity of shock wave calculated from the POS model fit of the QPO frequency evolution \citep[see,][]{DD13}. 
In Fig. 2a, we show variation of observed QPO 
frequencies with time. If viscosity parameter $\alpha$ were constant throughout the outburst then 
the variations of theoretically calculated QPO frequencies would be different. Dotted 
curve drawn for a viscosity parameter ($\alpha$) $0.001$ shows that
QPO frequency increases at almost constant rate. The 
dashed curve of Fig. 2a is for the effect of non-linear variation of the viscosity, which 
is shown in Fig. 2b. As the day progresses,  viscosity adjusts in such a way that the
angular momentum can produce a shock at a suitable place satisfying R-H conditions. 
\citet{CM95} and \citet{GC12}, 
with their extensive numerical simulations showed that angular momentum distribution 
depends on viscosity parameter. In our solution, at the beginning of the outburst 
during the hard state, from MJD=55417.2 to MJD=55420.2 (Debnath et al. 2013; MDC14), $\alpha$ varies 
from $1.3e$-$4$ to $5.9e$-$4$. During the hard-intermediate state, from MJD=55421.3 to MJD=55424.1 
\citep{DD13,MDC14}, 
$\alpha$ varies rapidly from $1.6e$-$3$ to $1.9e$-$2$. Our $\alpha$ calculation is for sub-Keplerian component only. 
In Fig. 2c, we show the variation of $\alpha$ with shock location. Dashed curve shows variation 
from our analytical solution, whereas dotted curve is a fitted polynomial, which gives a general trend
and could be used in other systems. We see that $\alpha \sim \cal K$ $X_s^{-q}$, 
where, $\cal K$(=$350.2$, with asymptotic standard error $6.29\%$) 
and $q$(=$2.34$, with asymptotic standard error $0.61\%$) are constants for this BHC.
It is to be noted that this viscosity parameter is computed for the sub-Keplerian flow component only.
 
\section{Discussions and Concluding Remarks}

QPOs in black hole candidates are very stable features. They are seen day after day, though the frequency
may be drifted slowly as the object goes from hard to soft state in the rising phase. This is generally 
observed in most of the outbursting BHCs 
\citep[][and references therein]{Nandi12,DD13}. 
Propagatory oscillating shock solution can explain such frequency drifts very well 
\citep{C08,DD13}. 
These phenomenological model is found to be justified when we actually compute shock drifts from radiated 
energy loss from a self-consistent transonic solution. We find that in order to have Rankine-Hugoniot conditions 
satisfied on each day, viscosity parameter must be evolving too. If the outer boundary condition is kept fixed, 
increase in viscosity parameter causes shock to drift outward 
\citep{CM95,GC12}, 
but if the inner boundary condition is kept fixed, the shock moves inward \citep{C90a,DC04}. 
We find support of the latter phenomenon in an outbursting source where matter supply is changing and viscosity 
enhancement steadily brought the shock closer to the black hole. Cooling is found to rise day by day and so is $\alpha$.
Such a movement of the shock increases QPO frequency as is observed. 
Our result establishes consistency in theoretical understanding of the observed data:
as cooling increases, observed QPO frequency increases due to drifting of the shock towards
black hole in a way that the cooling time scale roughly matches with the infall time scale.
This process brings the object towards the softer states as is observed. 
Shock locations were found to be located at the right place
(i.e., R-H conditions are satisfied), only if viscosity is not strictly constant, but gradually
rises from $0.0001$ to $0.02$ from the first day to $\sim$ seventh day. 
It is to be noted that there are alternative models \citep{TF04,TLM98}
where the corona is supposed to oscillate at its eigen frequency and the viscosity required in this 
case is around $0.1-0.5$. This appears to be too high as compared to what we find in the present paper. The
discrepancy could be due to the fact the latter models rely on oscillations of a Keplerian disk with 
high angular momentum and they require higher viscosity to reduce it drastically. In our case,
on the contrary, the oscillating CENBOL is highly sub-Keplerian to begin with. Therefore, a little 
viscosity is enough the transport requisite angular momentum.
This range of $\alpha$ we require is in the same ball park as 
obtained from numerical simulations \citep{Balbus91,Arlt01,Masada09} 
of magnetorotational instability (MRI).

\section*{Acknowledgments}

S. Mondal acknowledges the support of CSIR-NET scholarship.

\clearpage

\begin{figure}
\vspace {0.5cm}
\centering{
\includegraphics[scale=0.6,angle=0,width=10truecm]{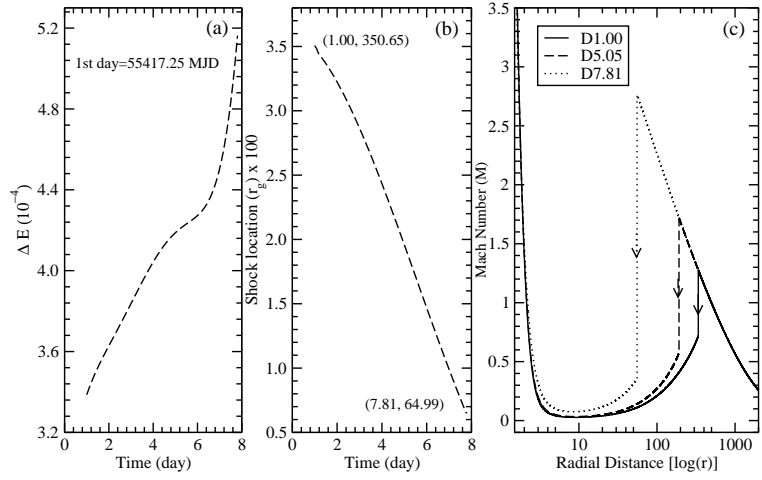}}
\caption{Variation of (a) cooling rate, and (b) shock location with time (in day) during rising phase of
H~1743-322 2010 outbutst are shown. In (c) mach number variation with logarithmic radial distance is shown.}
\label{fig1}
\end{figure}

\begin{figure}
\vspace {0.5cm}
\centering{
\includegraphics[scale=0.6,angle=0,width=10truecm]{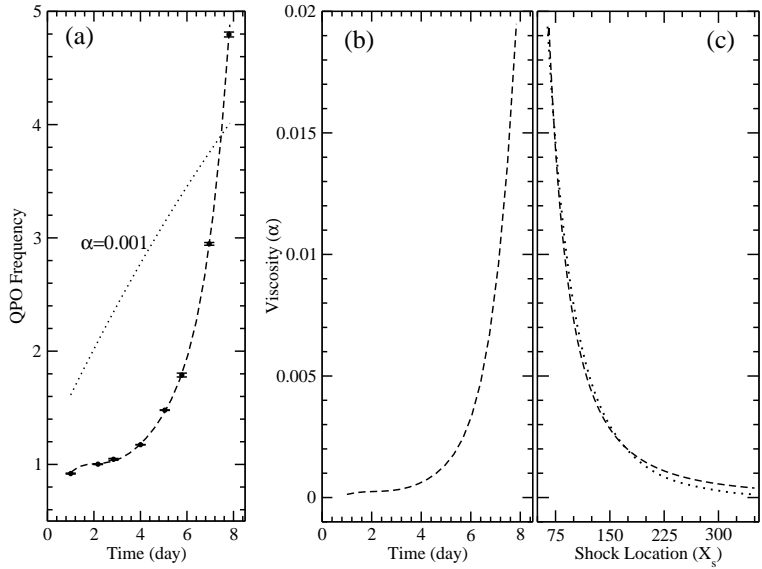}}
\caption{Variation of (a) QPO frequency with progressive days during the rising phase of
2010 outburst of H~1743-322, both obtained from observation and analytical solution. In (b) variation of viscosity 
with time (in day), and (c) with shock location are shown. Here viscosity is calculated only for the sub-Keplerian 
component of the accretion flow.}
\label{fig2}
\end{figure}

%
%

\end{document}